 \documentstyle[11pt]{article}
 
\def\D{{\cal D}}

\title{{\bf Toda Lattice and Tomimatsu-Sato Solutions}}

\author{{\sc Takeshi FUKUYAMA}$^\dagger$,\ %
       {\sc Kiyoshi Kamimura}$^\flat$  {\sc and}
        {\sc Songju YU}$^\dagger$\\
        \llap{$^\dagger$}%
        \small{\it{Department of Physics, Ritsumeikan University}}\\
        \small{\it{ Kusatsu, Shiga 525 JAPAN }}\\
        \llap{$^\flat$}%
        \small{\it{Department of Physics, Toho University}}\\
        \small{\it{Funabashi, 274 Japan}}\\
        {\it e-mails:} \small{FUKUYAMA\,@\,JPNYITP, KAMIMURA\,@\,JPNYITP}}

\date{}

\begin{document}

\maketitle

\thispagestyle{empty}

\begin{abstract}
We discuss an analytic proof of a conjecture (Nakamura) that
solutions of Toda molecule equation give those of Ernst equation
giving Tomimatsu-Sato solutions of Einstein equation. Using Pfaffian
identities it is shown for Weyl solutions completely and for generic
cases partially.
\vskip 20mm

\end{abstract}

\vfill
\vbox{
\hfill April 1995\null\par
\hfill Rits-TH-9511\null\par
\hfill TOHO-FP-9551}\null

\clearpage
Many of integrable non linear models are obtained from (anti) self-dual
Yang-Mills equations by reduction
\cite{ward}(Ward's conjecture).
Toda Lattice and Ernst equations are such examples of integrable systems.
However the relationship between these integrable systems is not so simple.
It is not affirmative that the system of Ernst equation belongs to the
conventional category of integrable systems. For instance, Lax pair for the
Ernst equation is not exactly the same type as those in typical soliton
theories\cite{papaharr}.

On the other hand, "{\it direct method}" developed  by  Hirota\cite{hiro}
has shown to be a very powerful tool in soliton theories.
One of the advantages of the direct method amongst others is that the proofs
of soliton solutions are reduced to Pfaffian identities.
The mathematical implication of the direct method was revealed by
Sato\cite{sato} .
Both the methods by Hirota and Sato seem to be quite useful in soliton
theories. However both theories have been seldom applied to the system of
Ernst equation except in very few articles.
In applying them to the system of Ernst equation
we expect to find the universal characters of soliton theories also in the
Ernst equation.
This is one of motivations of this article.

Recently Nakamura found an important relation between Tomimatsu-Sato
solutions (hereafter we refer it to TS) and Toda lattice solutions
\cite{anaka}.
He has shown that a series of TS solutions with  deformation parameters
$n$ are obtained as $n$-th lattice site amplitudes in a  special case of
general solutions of Toda molecule equation  (Nakamura's conjecture).
He has shown this proposition for small $n$ cases explicitly.
The aim of this article is to assure the conjecture analytically
and to take any step towards the scheme  mentioned above.
\vskip 6mm

We begin with discussing the Toda molecule equation describing a
semi-infinite lattice in two dimensions.
It is expressed in terms of the Hirota's bilinear forms \cite{hiro} as
\begin{equation}
  (D_X^2 - D_Y^2) \tau_n \cdot \tau_n - 2 \tau_{n + 1} \tau_{n - 1} = 0,
\label{eq:bitmeq1}
\end{equation}
where the Hirota derivative is defined by
$D(f\cdot g)=(\partial f)g-f(\partial g)$.
$n$'s are positive integers and the boundary condition is $\tau_0 = 1$
corresponding to the semi-infinite lattice.
The general solution of Eq.(\ref{eq:bitmeq1}) is expressed
in a form of two-directional Wronskian \cite{hos}
\begin{equation}
\tau_n = \det \pmatrix{
                   \psi     & L_- \psi     & \ldots & L_-^{n - 1} \psi
\cr                L_+ \psi & L_+ L_- \psi & \ldots & L_+ L_-^{n - 1} \psi
\cr                \vdots   & \vdots       & \vdots & \vdots
\cr        L_+^{n - 1} \psi & L_+^{n - 1} L_- \psi & \ldots &
   L_+^{n - 1}L_-^{n - 1} \psi \cr }.
\label{eq:tmsol}
\end{equation}

Here  $\psi$ is an arbitrary function and
$L_{\pm} \equiv \frac{\partial}{\partial X} \pm \frac{\partial}{\partial Y}$.
We first show how it appears as a result of a Pfaffian identity
for the help of later discussions, though it is a known fact.

We introduce $\D$ as a determinant of $(n+1)\times(n+1)$ matrix $\tau_{n+1}$;
\begin{equation}
  \D ~\equiv~\tau_{n+1} .  \label{eq:D}
\end{equation}

The minor $\D\biggl[{i \atop j}\biggl]$ is defined by deleting the $i$-th row
and the $j$-th column from $\D$.

Similarly $\D\biggl[{{i,k} \atop {j,l}}\biggl]$ is defined by deleting
the $i$ and $k$-th rows and the $j$ and $l$-th columns from $\D$ and so on.
The Toda molecule equation (\ref{eq:bitmeq1}) is now expressed as
\begin{equation}
  \D\biggl[{n \atop n}\biggl] \D\biggl[{{n + 1} \atop {n + 1}}\biggl] -
  \D\biggl[{{n + 1} \atop n}\biggl] \D\biggl[{n \atop {n + 1}}\biggl] -
  \D \D\biggl[{{n,n + 1} \atop {n,n + 1}}\biggl] = 0. \label{eq:tmji}
\end{equation}

It holds since it is nothing but the Jacobi's (Sylvester's) formula for
matrix minors, which is one of Pfaffian identities.
Thus Toda lattice equation has been reduced to Pfaffian identity
in direct method. It is also the case in the Einstein equation as will
be shown in the following.
\vskip 6mm

The Nakamura's conjecture on the Tomimatsu-Sato solutions consists of two
ingredients \cite{anaka};

(i) Ernst equation for axially symmetric metric of Einstein equation is
\begin{equation}
(\xi \xi^\ast - 1) \nabla^2 \xi - 2 \xi^\ast \nabla \xi \cdot \nabla \xi = 0.
\label{eq:erunsteq}
\end{equation}

Setting $~ \xi_n = \frac{g_n}{f_n}$,
Eq.(\ref{eq:erunsteq}) has a decomposition into two sets \cite{no}
\begin{eqnarray}
  & &D_x (g_n \cdot f_n - g_n^\ast \cdot f_n^\ast) = 0, \label{eq:tsdec1}\\
  & &D_y (g_n \cdot f_n + g_n^\ast \cdot f_n^\ast) = 0, \label{eq:tsdec2}
\end{eqnarray}
and
\begin{eqnarray}
  & &F (g_n^\ast \cdot f_n) = 0, \label{eq:tsdec3}\\
  & &F (g_n^\ast \cdot g_n + f_n^\ast \cdot f_n) = 0. \label{eq:tsdec4}
\end{eqnarray}

Here the bi-linear operator $F$ is
\begin{equation}
 F = (x^2 - 1) D_x^2 + 2 x \partial_x + (y^2 - 1) D_y^2 + 2 y \partial_y +
c_n \label{eq:F}
\end{equation}
and  $x$ and $y$ are usual prolated spheroidal coordinates.
\medskip

(ii) From the solutions of the Toda molecule equation $\tau_n$
a set of TS solutions are obtained as
\begin{equation}
  g_n~ =~\tau_n~=~\D\biggl[{{n + 1} \atop {n + 1}}\biggl], \hspace{1cm}
  f_n~=~\tau_{n-1}\mid_{_{\psi\rightarrow L_+L_-\psi}}~=~
\D\biggl[{{1,{n + 1}} \atop {1,{n + 1}}}\biggl] \label{eq:gnfn}
\end{equation}
by following choices of the arbitrary function $\psi$
and the constant $c_n$
\begin{equation}
  \psi = p x - i q y,~~~~~p^2~+~q^2~=~1,~~~~~c_n~=~-~2~n^2. \label{eq:psi}
\end{equation}

$x$ and $X$ are related by $\partial_X=(x^2-1)\partial_x$,~and~ $y$ and $Y$
are related by $\partial_Y=(y^2-1)\partial_y$.

It is very suggestive that $\xi_n = \frac{g_n}{f_n}$ in Eq.(\ref{eq:gnfn})
resembles with the solution $W_n$ of Sato equation\cite{sato}.
Our plan is to show the second statement (ii) analytically
making use of Pfaffian identities. The first set  Eqs.(\ref{eq:tsdec1}) and
(\ref{eq:tsdec2}) are proved in the generic case while the second set
Eqs.(\ref{eq:tsdec3}) and (\ref{eq:tsdec4}) are shown for a restricted case
of $q=0$.
\vskip 6mm

In order to prove the first set of equations  Eqs.(\ref{eq:tsdec1}) and
(\ref{eq:tsdec2}) we start to redefine $\D$ introduced in Eq.(\ref{eq:D}).
It is fit for treating  $f_n$, $g_n$ and their complex conjugate
$f_n^\ast$, $g_n^\ast$ on equal footing.
We first introduce complex conjugation operator $C$,
\begin{equation}
 C \psi = \psi^\ast = p x + i q y, \hspace{1cm} C^{2} = {\bf 1}.
 \label{eq:comop}\end{equation}

We define a new determinant $\D$ of $(n + 2) \times (n + 2)$ matrix as
follows,

\begin{equation}
\D = \det \pmatrix{
0         & C \psi     & L_+ C^2 \psi   & \ldots & L_+^n C^{n + 1} \psi
\cr
\psi   & L_+ C \psi & L_+^2 C^2 \psi & \ldots & L_+^{n + 1} C^{n + 1} \psi
  \cr
\vdots    & \vdots     &  \vdots        & \vdots & \vdots       \cr
L_+^n \psi & L_+^{n + 1} C \psi & L_+^{n + 2} C^2 \psi &
\ldots & L_+^{2 n + 1} C^{n + 1} \psi  \cr
                    }. \label{eq:newD}
                    \end{equation}

$g_n$ and $f_n$ in Eq.(\ref{eq:gnfn}) and their conjugate are expressed as
minors of $\D$;
\begin{equation}
  g_n = \D\biggl[{{1,{n + 2}} \atop {{n + 1},{n + 2}}}\biggl], \hspace{1cm}
    f_n = \D\biggl[{{1,2,{n + 2}} \atop {1,{n + 1},{n + 2}}}\biggl]
  \label{eq:newgnfn}
  \end{equation}
  and
  \begin{equation}
    g_n^\ast = \D\biggl[{{{n + 1},{n + 2}} \atop {1,{n + 2}}}\biggl],
    \hspace{10mm}
      f_n^\ast = \D\biggl[{{1,{n + 1},{n + 2}} \atop {1,2,{n + 2}}}\biggl].
   \label{eq:newgsnfsn}
   \end{equation}

 From the Wronskian structure of $\D$ in  Eq.(\ref{eq:newD}), the
derivatives of $f_n$ and $g_n$'s are also expressed as minors
\begin{eqnarray}
  & &L_+ g_n = \D\biggl[{{1,{n + 1}} \atop {{n + 1},{n + 2}}}\biggl],
  \hspace{1cm}
       L_+ f_n = \D\biggl[{{1,2,{n + 1}} \atop {1,{n + 1},{n + 2}}}\biggl],
     \nonumber \\
       & &L_+ g_n^\ast = \D\biggl[{{n,{n + 2}} \atop {1,{n + 2}}}\biggl],
  \hspace{17mm}
  L_+ f_n^\ast = \D\biggl[{{1,n,{n + 2}} \atop {1,2,{n + 2}}}\biggl]
  \label{eq:lpglpf}
  \end{eqnarray}
  and
  \begin{eqnarray}
 & &L_- g_n = \D\biggl[{{1,{n + 2}} \atop {n,{n + 2}}}\biggl], \hspace{18mm}
      L_- f_n = \D\biggl[{{1,2,{n + 2}} \atop {1,n,{n + 2}}}\biggl],
      \nonumber \\
 & &L_- g_n^\ast = \D\biggl[{{{n + 1},{n + 2}} \atop {1,{n + 1}}}\biggl],
  \hspace{1cm}
       L_- f_n^\ast = \D\biggl[{{1,{n + 1},{n + 2}} \atop {1,2,{n + 1}}}
       \biggl].
     \label{eq:lmglmf}
     \end{eqnarray}

We will prove that $f_n$ and $g_n$ given in Eq.(\ref{eq:gnfn})
satisfy Eqs.(\ref{eq:tsdec1}) and (\ref{eq:tsdec2}).
In showing it the explicit form of $\psi$ in Eq.(\ref{eq:psi}) is not
required.
In terms of $L_{\pm}$, Eq.(\ref{eq:tsdec1})~$\pm$~Eq.(\ref{eq:tsdec2}) are
given as
\begin{equation}
  g_n L_+ f_n - (L_+ g_n) f_n - g_n^\ast L_- f_n^\ast + (L_- g_n^\ast)
  f_n^\ast = 0, \label{eq:ts1pts2}
  \end{equation}
  \begin{equation}
    g_n L_- f_n - (L_- g_n) f_n - g_n^\ast L_+ f_n^\ast + (L_+ g_n^\ast)
  f_n^\ast = 0. \label{eq:ts1mts2}
  \end{equation}

Using Eqs.(\ref{eq:lpglpf}) and (\ref{eq:lmglmf}),
Eq.(\ref{eq:ts1pts2}) becomes
\begin{eqnarray}
 & & \D\biggl[{{1,n + 2} \atop {n + 1,n + 2}}\biggl] \D\biggl[{{1,2,n + 1}
 \atop {1,n + 1,n + 2}}\biggl] \nonumber \\
  &-& \D\biggl[{{1,n + 1} \atop {n + 1,n + 2}}\biggl] \D\biggl[{{1,2,n + 2}
 \atop {1,n + 1,n + 2}}\biggl] \nonumber \\
  &-& \D\biggl[{{n + 1,n + 2} \atop {1,n + 2}}\biggl]
  \D\biggl[{{1,n + 1,n + 2}  \atop {1,2,n + 1}}\biggl] \nonumber \\
   &+& \D\biggl[{{n + 1,n + 2} \atop {1,n + 1}}\biggl]
   \D\biggl[{{1,n + 1,n + 2}
 \atop {1,2,n + 2}}\biggl]
  = 0. \label{eq:Dts1pts2}
  \end{eqnarray}

Eq.(\ref{eq:Dts1pts2}) holds using a following Pfaffian identity.
For any determinant $A$ of an arbitrary square matrix
\begin{equation}
A\biggl[{{a,b} \atop {d,e}}\biggl]A\biggl[{{c} \atop {e}}\biggl] +
A\biggl[{{b,c} \atop {d,e}}\biggl]A\biggl[{{a} \atop {e}}\biggl] +
A\biggl[{{c,a} \atop {d,e}}\biggl]A\biggl[{{b} \atop {e}}\biggl] = 0,
\label{eq:newmatid}
\end{equation}
where $a>b>c$.
Leaving the proof of this identity later on we accept it.

Setting $A = \D\biggl[{1 \atop {n + 2}}\biggl]$, the first two terms
of  Eq.(\ref{eq:Dts1pts2}) become
$$ A\biggl[{{2,n + 1} \atop {1,n + 1}}\biggl]A\biggl[{{n + 2} \atop
  {n + 1}}\biggl] - A\biggl[{{2,n + 2} \atop {1,n + 1}}\biggl]A
  \biggl[{{n + 1} \atop {n + 1}}\biggl]~=~
 - A\biggl[{{n + 2,n + 2} \atop {1,n + 1}}\biggl]A
  \biggl[{{2} \atop {n + 1}}\biggl]
$$
\begin{eqnarray}
    = - \D\biggl[{{1,2} \atop {n + 1,n + 2}}\biggl]
    \D\biggl[{{1,n + 1,n + 2}
  \atop {1,n + 1,n + 2}}\biggl].
  \end{eqnarray}

Similarly the last two terms of  Eq.(\ref{eq:Dts1pts2}) are, by
setting $A = \D\biggl[{{n + 2} \atop 1}\biggl]$, as
\begin{equation}
   \D\biggl[{{n + 1,n + 2} \atop {1,2}}\biggl]\D\biggl[{{1,n + 1,n + 2}
  \atop {1,n + 1,n + 2}}\biggl].
  \end{equation}

The sum of above two equations vanishes since
  \begin{equation}
     \D\biggl[{{1,2} \atop {n + 1,n + 2}}\biggl] =
     \D\biggl[{{n + 1,n + 2} \atop
   {1,2}}\biggl],
\end{equation}
which is one of the properties of $\D$ in Eq.(\ref{eq:newD}).
Eq.(\ref{eq:ts1mts2}) is also proved analogously to the above.
\vskip 6mm

The crucial identity  Eq.(\ref{eq:newmatid}) we have used is proved as
follows.
Let us consider a determinant of $n \times n$ anti-symmetric matrix $A$ whose
$(i,j)$ element is $a_{ij}(=-a_{ji})$. Pfaffian $(1,2,\ldots,n)$ is defined by
\begin{equation}
 det A \equiv \left\{
                   \begin{array}{ll}
                       (1,2,\ldots,n)^2 & \mbox{for even $n$} \\
                       0 & \mbox{for odd $n$}
                   \end{array}
                \right.
\end{equation}
with $a_{ij}=(i,j)$.

By definition Pfaffian is expanded as
\begin{equation}
(1,2,\ldots,n) = \sum_{j = 1}^n (-)^j (1,j) (2,3,\ldots,\hat{j},\ldots,n)
\end{equation}
where hatted component is removed one.

 We can also express a determinant of $(n + 2)$-th order matrix as
Pfaffian of $2(n+2)$-th order by
\begin{eqnarray}
 det~a_{ij} = (1,2,\ldots,n+1,n+2,n+2^\ast,n+1^\ast,\ldots,2^\ast,1^\ast)
 \nonumber \\
  (i,j = 1,2,\ldots,n+2)
\end{eqnarray}
supplemented by the condition,
$(i,j) = (i^\ast,j^\ast) = 0$, $(i,j^\ast) \equiv a_{ij}$.

  From the antisymmetry property of Pfaffians it follows
\begin{equation}
   det~a_{ij} = (-)^P (a,b,c,1,\ldots,\hat{a},\ldots,\hat{b},\ldots,
   \hat{c},
 \ldots,n+2,n+2^\ast,\ldots,\hat{d}^\ast,\ldots,\hat{e}^\ast,\ldots,
 1^\ast,
 d^\ast,e^\ast),
 \end{equation}
 where
$(-)^P$ is $\pm 1$ depending on the signature of permutation $P$ of
reordering.
In terms of Pfaffians each terms of Eq.(\ref{eq:newmatid}) are expressed
as follows.
\begin{eqnarray}
 & & A\biggl[{{a,b} \atop {d,e}}\biggl] = (c,1,\ldots,\hat{a},
 \ldots,\hat{b},
 \ldots,\hat{c},\ldots,\hat{d}^\ast,\ldots,\hat{e}^\ast,\ldots,2^
 \ast,1^\ast)
 \nonumber \\
  & & A\biggl[{{c} \atop {e}}\biggl] =  (a,b,1,\ldots,\hat{a},
  \ldots,\hat{b},
 \ldots,\hat{c},\ldots,\hat{d}^\ast,\ldots,\hat{e}^\ast,\ldots,2^
 \ast,1^\ast,
 d^\ast) \nonumber \\
  & & A\biggl[{{b,c} \atop {d,e}}\biggl] = (a,1,\ldots,\hat{a},
  \ldots,\hat{b},
 \ldots,\hat{c},\ldots,\hat{d}^\ast,\ldots,\hat{e}^\ast,
 \ldots,2^\ast,1^\ast)
 \nonumber \\
  & & A\biggl[{{a} \atop {e}}\biggl] = (b,c,1,\ldots,\hat{a},
  \ldots,\hat{b},
 \ldots,\hat{c},\ldots,\hat{d}^\ast,\ldots,\hat{e}^\ast,
 \ldots,2^\ast,1^\ast,
 d^\ast) \\
  & & A\biggl[{{a,c} \atop {d,e}}\biggl] = (b,1,\ldots,\hat{a},
 \ldots,\hat{b},
 \ldots,\hat{c},\ldots,\hat{d}^\ast,\ldots,\hat{e}^\ast,
 \ldots,2^\ast,1^\ast)
 \nonumber \\
  & & A\biggl[{{b} \atop {e}}\biggl] = (a,c,1,\ldots,\hat{a},
 \ldots,\hat{b},
 \ldots,\hat{c},\ldots,\hat{d}^\ast,\ldots,\hat{e}^\ast,
 \ldots,2^\ast,1^\ast,
 d^\ast) \nonumber \label{eq:teigi}
 \end{eqnarray}
 up to common signature factor $(-)^P$
provided that the order of $a, b$ and $c$ is kept in mind as in
Eq.(\ref{eq:newmatid}).
Eq.(\ref{eq:newmatid}) is expressed symbolically in terms of Maya diagram
\cite{sato} as
\begin{eqnarray}
 & &A\biggl[{{a,b} \atop {d,e}}\biggl]A\biggl[{{c} \atop {e}}\biggl] +
 A\biggl[{{b,c} \atop {d,e}}\biggl]A\biggl[{{a} \atop {e}}\biggl] -
 A\biggl[{{a,c} \atop {d,e}}\biggl]A\biggl[{{b} \atop {e}}\biggl]
 \nonumber \\
  &=& \stackrel{a}{\fbox{~X~}}\stackrel{b}{\fbox{~X~}}\stackrel{c}
  {\fbox{~O~}}
 \stackrel{d^\ast}{\fbox{~X~}} \times \stackrel{a}{\fbox{~O~}}
 \stackrel{b}
 {\fbox{~O~}}\stackrel{c}{\fbox{~X~}}\stackrel{d^\ast}{\fbox{~O~}}
 \nonumber \\
  &+& \fbox{~O~}\fbox{~X~}\fbox{~X~}\fbox{~X~} \times \fbox{~X~}\fbox{~O~}
 \fbox{~O~}\fbox{~O~} \nonumber \\
  &-& \fbox{~X~}\fbox{~O~}\fbox{~X~}\fbox{~X~} \times \fbox{~O~}\fbox{~X~}
 \fbox{~O~}\fbox{~O~} \nonumber \\
  &=& \fbox{~O~}\fbox{~O~}\fbox{~O~}
 \fbox{~X~} \times \fbox{~X~}\fbox{~X~}\fbox{~X~}\fbox{~O~}.
 \end{eqnarray}

The last equality holds by virtue of Pfaffian identity.
Both $\fbox{~O~}\fbox{~O~}\fbox{~O~}\fbox{~X~}$ and $\fbox{~X~}
\fbox{~X~}\fbox{~X~}\fbox{~O~}$ consist of different numbers with and
without asterisks and vanish by their definition.
It completes the proof of the identity Eq.(\ref{eq:newmatid}).
\vskip 6mm

Next we discuss
the second set of decomposition equations Eqs.(\ref{eq:tsdec3})
and (\ref{eq:tsdec4}). In contrast to the previous case the explicit
form of $\psi$ in Eq.(\ref{eq:psi}) is required.
$g_n(f_n)$ is the determinant of matrix whose $(i,j)
\biggl((i-1,j-1)\biggr)$
element is
\begin{eqnarray}
 L_+^{i-1} L_-^{j-1} \psi &=& L_+^{i-1} L_-^{j-1} (p x - i q y) \nonumber
 \\                       &=& p W_{i+j-1}(x) + (-1)^j i q W_{i+j-1}(y),
                          \end{eqnarray}
                          where
                          \begin{equation}
                           W_{n + 1}(z) = (z^2 - 1) \frac{d}{dz} W_{n}(z)
                           \hspace{1cm} with \hspace{1cm}
 W_1(z) = z. \label{eq:W}
 \label{eq:gfij}
 \end{equation}

In this paper we restrict ourselves to prove for a case of $q = 0$
and the general case is left for future.
In case of $p=1$ and $q=0$,~ $\psi = x$  and
$g_n$ and $f_n$ are real functions depending only on $x$.
Explicit forms of $g_n$ and $f_n$ are
\begin{equation}
  g_n = \det \pmatrix{
                     W_1      & W_2      & \ldots  & W_n        \cr
                   W_2      & W_3      & \ldots  & W_{n+1}    \cr
                   \vdots   & \vdots   & \vdots  & \vdots     \cr
                   W_n      & W_{n+1}  & \ldots  & W_{2n-1}   \cr
                                      },~~~~~
  f_n = \det \pmatrix{
                          W_3      & \ldots  & W_{n+1}    \cr
                          \vdots   & \vdots  & \vdots     \cr
                          W_{n+1}  & \ldots  & W_{2n-1}   \cr
                                      }. \label{eq:gn}
                                      \end{equation}

Using Eq.(\ref{eq:gfij}) we can evaluate the determinant,
\begin{eqnarray}
  g_n &=& \det \pmatrix{
    x          & 1          & 0       & 0    & \ldots    & 0   \cr
 x^2 - 1       & 2 x        & 2       & 0    & \ldots    & 0   \cr
2 x (x^2 - 1)  & 6 x^2 -1   & 12 x    & 12   & \ldots    & 0   \cr
\vdots         & \vdots & \vdots  & \vdots  & \ldots & \vdots  \cr
                                      } \nonumber\\
&=&  \frac{A_n}{2}(x^2 -1)^{\frac{n (n - 1)}{2}}
\biggl((x + 1)^n + (x - 1)^n\biggr), \label{eq:gnx}
      \end{eqnarray}
where the coefficient $A_n$ is
\begin{equation}
A_n = (n - 1)^2 (n - 2)^4 \ldots 2^{2 (n - 1)}.\label{eq:An}
      \end{equation}

The equation  (\ref{eq:gnx}) is proved by induction.
For $n = 1$ and $n=2$ cases Eq.(\ref{eq:gnx}) holds.
Assuming it for $n = l-1$ and $n=l$ cases
we prove it for the case of $n = l + 1$.
Applying Jacobi's formula (\ref{eq:tmji}) to $g_n$ we obtain,
corresponding to the Toda molecule equation in Eq.(\ref{eq:tmsol}),
 \begin{equation}
    g_{l - 1}~g_{l + 1} = (L_X^2~ g_l)~g_l - (L_X ~g_l)^2, \label{eq:gjaco}
    \end{equation}
 where $L_X \equiv (x^2 - 1) \partial_x$.
 Using the assumed forms for $g_l$ and $g_{l-1}$ we find the expected form for
$g_{l+1}$;
  \begin{equation}
   g_{l + 1} =  \frac{A_{l + 1}}{2} (x^2 -1)^{\frac{(l + 1) l}{2}}
 \biggl((x + 1)^{l + 1} + (x - 1)^{l + 1}\biggr).
 \end{equation}

 Here we have used an equality obtained from the definition of $A_n$
in Eq.(\ref{eq:An});
 \begin{equation}
  A_{n - 1} A_{n + 1} = n^2 A_n.
  \end{equation}

Thus Eq.(\ref{eq:gnx}) is proved for $n = l + 1$.
Quite analogously $f_n$ is shown to be
\begin{equation}
 f_n =  \frac{A_n}{2}(x^2 -1)^{\frac{n (n - 1)}{2}}
\biggl((x + 1)^n -  (x - 1)^n\biggr). \label{eq:fnx}
 \end{equation}

The second set of the decoupled equations
Eqs.(\ref{eq:tsdec3}) and (\ref{eq:tsdec4}) are
 \begin{eqnarray}
  & &F (g_n \cdot f_n) = 0, \nonumber \\
   & &F (g_n \cdot g_n + f_n \cdot f_n) = 0, \label{eq:wdec34}
   \end{eqnarray}
and $F$ becomes in $q=0$ and $c_n=-2n^2$ case
$$ F a \cdot b = \frac{1}{x^2 - 1} \biggl((L_X^2 a) b + a (L_X^2 b) -
2 (L_X a) (L_X b)\biggr) - 2 n^2 a b.$$

It is straight forward to show $g_n$ and $f_n$ in Eqs.(\ref{eq:gnx}) and
(\ref{eq:fnx}) satisfy Eq.(\ref{eq:wdec34}).
\vskip 6mm

In this paper we have discussed a proof of Nakamura's conjecture on
Tomimatsu-Sato solutions. In the proof a set of Pfaffian identities played
crucial roles. The first set of decoupled equations for Ernst equation
Eqs.(\ref{eq:tsdec1}) and (\ref{eq:tsdec2}) has been shown analytically.
Here the explicit form of $\psi$ in Eq.(\ref{eq:psi}) is not required.
The second set of decoupled equations Eqs.(\ref{eq:tsdec3}) and
(\ref{eq:tsdec4}) has also proved for the restricted case of $q=0$.
They are corresponding to the (extended)  Weyl solutions.
There remains to show the most general case, $q\not=0$.
We expect that Pfaffian identities will play important roles
in the analytic proof as in the case shown here.
It is left as future problem.


\end{document}